# A transfer-learning-enhanced POD-FNN surrogate for rapid signal prediction and inverse fitting in thermoreflectance with patterned transducers

Bingjia Xiao[1], Tao Chen[1], Puqing Jiang[1]

[1]*School of Energy and Power Engineering, Huazhong University of Science and Technology, Wuhan, Hubei 430074, China*

**ABSTRACT:** Patterned-transducer thermoreflectance can enhance sensitivity to low-thermal-conductivity materials by suppressing lateral heat spreading in the metal transducer, but its practical use is still limited by the cumulative cost of repeated high-fidelity forward evaluations required in iterative fitting and related analyses. In this work, a transfer-learning-enhanced POD-FNN surrogate is developed for rapid phase-signal prediction in patterned-transducer thermoreflectance, using patterned-transducer FDTR as a representative platform. A high-fidelity COMSOL model is first established and validated, and proper orthogonal decomposition is then applied directly to the phase signals to construct a compact reduced-order representation. Based on this representation, a feedforward neural network is trained to predict the corresponding POD coefficients from thermophysical and geometric parameters. Within the original parameter domain, the surrogate achieves mean and median RMSE values of $0.19°$ and $0.17°$, respectively, with the maximum RMSE below $0.47°$, while reducing the average prediction time per signal from 5.39 s for COMSOL to 0.01 s, corresponding to a speedup of about 534×. In surrogate-based inverse analysis, the fitting time for a representative case is reduced from about 18950 s to about 65 s while maintaining comparable accuracy. The framework is further applied to measured $Al/SiO_2$ samples, for which the extracted silica thermal conductivity remains stable at $1.44 \pm 0.088$, $1.43 \pm 0.093$, and $1.50 \pm 0.079$ W/(m·K) for conventional FDTR and patterned FDTR with $R_{pat} = 5.3$ μm and $3.25$ μm, respectively. Transfer learning is then introduced to extend the surrogate to expanded parameter domains. Compared with the source-domain baseline, the transfer-learned surrogates further reduce the prediction error, with transfer learning on full-range expanded target dataset (TL-FR) giving the best overall performance. At the same time, reducing the additional target-domain dataset from 6000 to 1000 samples lowers the high-fidelity data-generation time from about 34179 s to about 5885 s. These results demonstrate that the proposed framework provides an accurate and computationally efficient route for patterned thermoreflectance workflows requiring repeated forward evaluation, rapid inverse fitting, and cost-effective model updating across expanded parameter domains. From an application perspective, the proposed framework is attractive for high-throughput thermal-property screening, mapping-oriented characterization, and rapid model updating in thin-film and interface-dominated thermal systems relevant to thermal management and device engineering.

**Keywords:** Patterned-transducer thermoreflectance; Frequency domain thermoreflectance (FDTR); Surrogate modeling; Proper orthogonal decomposition; Neural network; Transfer learning

## I. Introduction

Thermoreflectance techniques provide a non-contact means to probe heat transport in thin films, multilayer structures, and buried interfaces while retaining high temporal resolution and micrometer-scale spatial sensitivity[1-3]. Among them, frequency-domain thermoreflectance (FDTR) has become one of the most widely used approaches because it employs a relatively simple optical configuration and links material properties to the frequency response of the surface temperature. In a typical FDTR measurement, a thin metal transducer deposited on the sample surface absorbs the modulated pump beam and simultaneously serves as the optical thermometer through the thermoreflectance effect[4].

Conventional FDTR, however, faces a clear limitation when applied to low-thermal-conductivity materials. A significant portion of the deposited heat can spread laterally within the metal transducer before entering the sample, which weakens sensitivity to the substrate response and reduces the robustness of parameter extraction[5-7]. Consequently, quantitative analysis of the present configuration relies primarily on numerical forward modeling. To address this issue, Akura et al. patterned the transducer in FDTR and showed through finite-element analysis and experiments that restricting lateral heat flow within the transducer improves the precision of thermal conductivity measurements for low-conductivity materials[8]. Warzoha et al. explored a related concept by introducing microscale confinement into FDTR to probe thermal resistance across buried interfaces. Their study showed that radial confinement suppresses heat spreading in the upper layers, increases the effective thermal penetration depth, and enhances sensitivity to interfaces located well below the surface[9]. Together, these studies show that patterned thermoreflectance does more than modify the measurement geometry; it reshapes the heat-flow pathway and can improve sensitivity to targeted thermal parameters.

At the same time, the finite lateral confinement introduced by patterning makes forward modeling more challenging. Standard FDTR fitting usually relies on analytical or semi-analytical heat-transfer models developed for laterally infinite or semi-infinite multilayer systems. Once the transducer is patterned and finite lateral boundaries become important, those assumptions no longer hold. Recent studies on patterned or confined FDTR have therefore relied on finite-element simulations for forward signal generation and quantitative fitting [2,3][10]. Although semi-analytical treatments have been developed for some special laterally constrained geometries, such as finite particles, no generally applicable analytical or semi-analytical solution is available for the patterned-transducer multilayer geometry considered here. Consequently, quantitative analysis of the present configuration relies primarily on numerical forward modeling. This approach provides the flexibility required for realistic geometries, but it becomes computationally demanding when the model must be solved repeatedly for iterative parameter estimation and sensitivity-based analysis. Although the cost of a single high-fidelity evaluation may be acceptable, the cumulative cost can become substantial in patterned thermoreflectance when the finite-element forward model is invoked many times during fitting and related analyses.

Reduced-order modeling offers a practical route to alleviate this burden. Proper orthogonal decomposition (POD) compresses high-dimensional simulation data into a small set of dominant modes while preserving the main structure of the original solution[11-13]. Tang et al. used a POD-based reduced-order model to rapidly predict velocity and temperature in data centers and demonstrated that reduced-order methods can achieve both good accuracy and substantial speedup in complex thermal-fluid systems[14]. Li et al. developed a data-driven reduced-order model for gas-solid heat transfer in fluidized beds and further illustrated the value of reduced-order techniques for fast thermal prediction[15]. Xiang et al. later combined POD, neural networks, and transfer learning for thermal-field prediction in high-heat-flux electronic systems, showing that reduced-order representations can work effectively with machine learning to improve both prediction efficiency and model adaptability[16]. These advances make POD-based surrogate modeling well-suited for patterned thermoreflectance[17], where the focus is on thermoreflectance signal generated by a computationally expensive numerical solver rather than a full temperature field.

A reduced surrogate, however, does not fully resolve the problem on its own. Its predictive accuracy usually depends on the parameter domain covered by the training set, and performance often degrades once the target conditions move beyond that original range[18, 19]. Reconstructing the surrogate over an expanded domain typically requires many new high-fidelity simulations, which can offset much of the computational advantage gained from reduced-order modeling. Transfer learning offers a more economical alternative because it allows a pretrained model to retain previously learned features and adapt to a related target domain with limited additional data[20-22]. In thermal-system optimization, Zhang et al. showed that transfer learning can combine abundant low-fidelity information with a small number of high-fidelity simulations to improve prediction accuracy while substantially reducing the demand for expensive high-fidelity samples[23]. Patterned thermoreflectance also provides a favorable setting for transfer learning, because thermoreflectance signals are often governed by a limited number of coupled parameter combinations, and their phase responses tend to vary smoothly over neighboring thermophysical and geometric conditions[24].

In this work, we develop a transfer-learning-enhanced POD-FNN surrogate for rapid signal prediction in patterned thermoreflectance, using patterned FDTR as a representative platform. The main contributions of this study are threefold. First, a reduced-order surrogate framework is established for patterned thermoreflectance by combining a high-fidelity finite-element forward model with POD-based output compression and neural-network-based coefficient prediction. Second, POD is applied directly to frequency-dependent phase signals rather than to the full thermal field, which better matches the signal-level tasks of forward prediction and inverse fitting. Third, transfer learning is incorporated to extend the surrogate to expanded parameter domains with limited additional high-fidelity data. The proposed framework is validated through forward-model comparison, surrogate performance assessment, synthetic inverse analysis, measured-signal fitting, and transfer-learning evaluation. More importantly, the present reduced-order strategy is constructed directly at the signal level rather than at the thermal-field level. This distinction is not merely technical. In patterned thermoreflectance, both forward evaluation and inverse fitting are ultimately performed on the measured phase signal, whereas the full temperature field is only an intermediate quantity in the high-fidelity solver. Therefore, direct reduction of the phase response provides a representation that is more tightly aligned with the actual characterization task and more suitable for repeated signal prediction in iterative fitting workflows. From an engineering perspective, such a capability is valuable not only for accelerating patterned FDTR signal analysis itself, but also for enabling faster thermal characterization workflows for thin films, interfaces, and low-thermal-conductivity materials relevant to thermal management and advanced device design[25]. In this sense, the proposed framework can be viewed as a surrogate-assisted thermal metrology tool that supports more efficient use of high-fidelity modeling in engineering characterization tasks. Beyond accelerating standalone signal prediction, the proposed framework is also relevant to practical thermal-characterization workflows in which a large number of forward evaluations are required across multiple measurement locations or repeated sample conditions. This includes, for example, high-throughput spatial mapping of effective thermal properties[26], rapid screening of low-thermal-conductivity samples and buried interfaces[10], and efficient framework recalibration during device development and manufacturing. The detailed methodology is presented in Section 2, followed by the experimental and numerical settings in Section 3 and the results and discussion in Section 4.

**II. Methodology**

### 2.1 Proper orthogonal decomposition

In this work, POD is used to construct a reduced-order representation of the thermoreflectance signals generated by the high-fidelity forward model. Unlike many conventional POD applications, where the quantity to be reduced is a temperature or flow field, the present study applies POD directly to the frequency-dependent phase signal. This treatment is more consistent with the objective of the present work, since both forward prediction and parameter fitting are ultimately performed on the phase response rather than on the full thermal field. The overall POD procedure is illustrated in Fig. 1, including the construction of signal snapshots, the cumulative energy criterion for mode truncation, and the retained POD modes.

Each sample in the parameter space is described by a vector of thermophysical and geometric variables,

$$\mu = [\mu_1, \mu_2, \ldots, \mu_p]^T, \qquad (1)$$

and the corresponding phase signal sampled at $N_f$ modulation frequencies is denoted by $\phi(\mu) \in R^{N_f}$. After collecting the signals from $N_s$ full-order simulations, the snapshot matrix can be assembled as

$$\mathbf{\Phi} = [\phi^{(1)}, \phi^{(2)}, \ldots, \phi^{(N_s)}] \in R^{N_f \times N_s}, \qquad (2)$$

where $\phi^{(i)} = \phi(\mu^{(i)})$ is the phase signal corresponding to the $i$-th parameter sample. Representative phase snapshots that form the snapshot matrix are shown in Fig. 1(a). Before performing POD, the mean signal is removed from the snapshot ensemble. The mean signal is given by

$$\bar{\phi} = \frac{1}{N_s} \sum_{i=1}^{N_s} \phi^{(i)}, \tag{3}$$

and the centered snapshot matrix is therefore written as

$$\mathbf{\Phi}_c = \left[\phi^{(1)} - \bar{\phi}, \phi^{(2)} - \bar{\phi}, \ldots, \phi^{(N_s)} - \bar{\phi}\right]. \tag{4}$$

This step allows the resulting modes to describe the dominant variations among different phase signals rather than the common baseline shared by all samples. Singular value decomposition is then applied to the centered snapshot matrix: $\mathbf{\Phi}_c = \mathbf{U\Sigma V}^T$, where the columns of $\mathbf{U}$ form an orthonormal basis, and the diagonal entries of $\mathbf{\Sigma}$ are the singular values arranged in descending order. The leading modes capture the dominant structures of the signal ensemble, whereas the contribution of higher-order modes gradually decreases. To obtain a compact reduced basis, only the first r dominant modes are retained,

$$\mathbf{M}_r = [M_1, M_2, \ldots, M_r] \in R^{N_f \times r}, \tag{5}$$

where $r \ll N_f$. The retained dimension is determined according to the cumulative energy ratio,

$$\eta(r) = \frac{\sum_{j=1}^{r} \sigma_j^2}{\sum_{j=1}^{m} \sigma_j^2}, \tag{6}$$

where $\sigma_j$ is the $j$-th singular value and $m = \min(N_f, N_s)$. Fig. 1(b) shows the cumulative energy ratio as a function of retained mode number, and Fig. 1(c) shows several retained POD modes. In practice, $r$ is chosen as the smallest value that satisfies a prescribed energy threshold (the energy threshold is set to 99.9999% in this work), so that the main characteristics of the full-order phase signals are preserved while the output dimension is substantially reduced. In the present study, the cumulative energy ratio reached 99.99994% with 15 retained modes. Therefore, $r = 15$ was adopted for subsequent POD reconstruction and surrogate training. A comparison of phase-signal accuracy obtained under the 99.999% and 99.9999% energy thresholds is provided in the Supplementary Material. With this reduced basis, the reconstructed phase signal can be written as

$$\hat{\phi}(\mu) = \bar{\phi} + \mathbf{M}_r \cdot a(\mu), \tag{7}$$

where

$$a(\mu) = [a_1(\mu), a_2(\mu), \ldots, a_r(\mu)]^{\mathrm{T}} \tag{8}$$

is the vector of POD coefficients. For a known full-order signal, the corresponding coefficients are obtained by orthogonal projection,

$$a(\mu) = \mathbf{M}_r^{\mathrm{T}}(\phi(\mu) - \bar{\phi}). \tag{9}$$

After this transformation, the original prediction problem is no longer expressed in terms of the full frequency-response vector, but in terms of a much smaller set of POD coefficients. This reduction is well suited to the present problem because, although the phase signal spans many frequency points, its overall shape remains highly correlated across the parameter space. Therefore, a relatively small number of POD modes can capture most signal variations. POD eliminates redundancy in the output space and makes it easier for the neural-network surrogate to learn the mapping from the input parameter vector $\mu$ to the reduced-order coefficient vector $a(\mu)$.

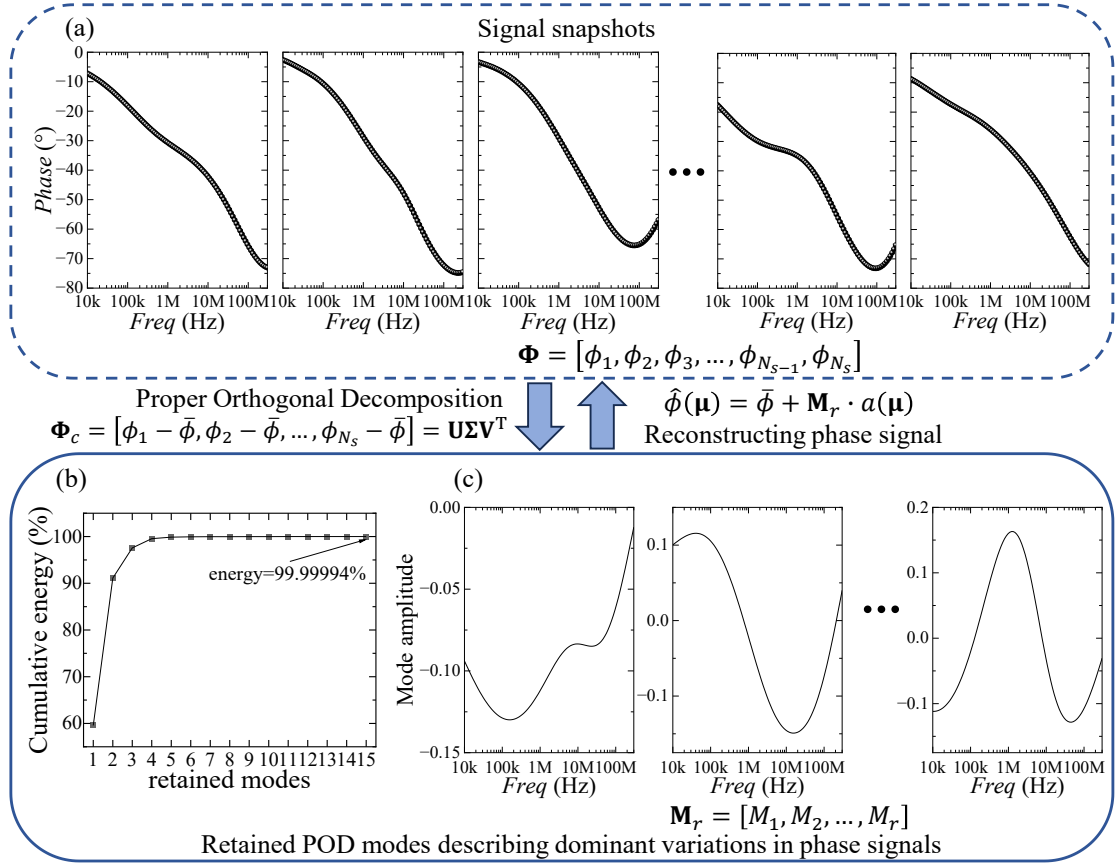

**Fig. 1.** POD dimensionality reduction of FDTR phase signals. (a) Phase signal snapshots. (b) Cumulative energy ratio versus retained modes. (c) Retained POD modes.

### 2.2 Neural-network surrogate model

After POD reduces the full-order phase signal to a low-dimensional coefficient vector, the remaining task is to establish the mapping from the input parameter vector $\mu$ to the reduced-order coefficient vector $a(\mu)$. In this work, this nonlinear mapping is learned by a feedforward neural network (FNN), which serves as the surrogate model for rapid phase prediction. The architecture of the POD-FNN surrogate is illustrated in Fig. 2.

The input of the network is the parameter vector $\mu$, which contains the thermophysical and geometric variables defining each sample, whereas the output is the corresponding POD coefficient vector $a(\mu)$. Since the input parameters span different physical dimensions and numerical ranges, directly using their raw values is not favorable for network training. In the present work, the input parameters are used in their original form, and both the input variables and the output POD coefficients are linearly normalized to the interval [0,1] before network training. The same preprocessing and normalization procedures are also applied during prediction.

A feedforward neural network is adopted to approximate the relationship between the input parameters and the reduced-order coefficients. In the present study, the network contains two hidden layers with 32 neurons in each layer. This compact architecture is sufficient to capture the nonlinear mapping from the input parameters to the reduced-order coefficients, while avoiding unnecessary model complexity for the present low-dimensional regression task after POD reduction. The samples are randomly divided into training, validation, and test subsets with ratios of 0.8, 0.1, and 0.1, respectively, using MATLAB's dividerand function. The network is trained by Bayesian regularization backpropagation (trainbr), which provides stable training and good generalization for regression problems of the present scale and does not rely strongly on validation-based early stopping.

For a new input sample $\mu$, the trained network outputs the predicted coefficient vector

$$\hat{a}(\mu) = [\hat{a}_1(\mu), \hat{a}_2(\mu), \ldots, \hat{a}_r(\mu)]^T. \tag{10}$$

After inverse normalization, the predicted coefficients are substituted into Eq. (7) to reconstruct the phase signal. In this way, the original problem of predicting the full frequency-response curve is converted into the prediction of a small number of POD coefficients followed by an inexpensive reconstruction step. Once training is completed, online prediction only requires a forward pass through the neural network and the subsequent POD reconstruction, which is much more efficient than repeatedly calling the high-fidelity numerical solver.

This POD-FNN framework is well suited to the present problem. Although the phase signal varies with thermophysical properties, geometric parameters, and measurement conditions, its variation across the parameter space remains strongly correlated, which allows a small number of POD modes to capture the dominant response features. The neural network then learns the nonlinear dependence of these reduced-order coefficients on the input parameters, thereby providing an efficient surrogate for rapid forward prediction of patterned thermoreflectance signals.

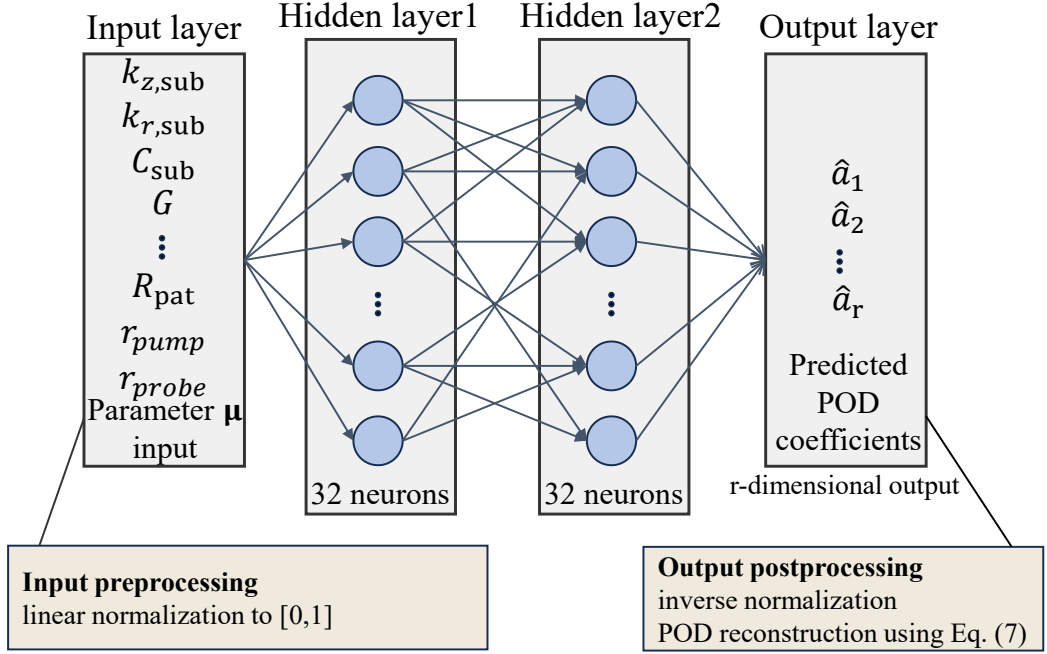

**Fig. 2.** Schematic architecture of the POD-FNN surrogate model.

### 2.3 Transfer learning

Although the POD-FNN surrogate predicts the phase response accurately within the original parameter domain, its performance can degrade after the parameter range is expanded. A straightforward solution is to regenerate training samples in the expanded domain and retrain the surrogate from scratch. However, such a strategy requires a substantial number of additional high-fidelity simulations, which undermines the computational efficiency of the reduced-order framework. To avoid this cost, we introduce transfer learning to adapt the pretrained source-domain surrogate to the expanded target domain. The overall transfer-learning strategy is schematically illustrated in Fig. 3.

In this work, we treat the surrogate trained over the original parameter range as the source model and the surrogate associated with the expanded parameter range as the target model. Instead of constructing a new surrogate from the beginning, we use the pretrained source-domain network to initialize the target-domain model and then fine-tune it using target-domain samples together with a retained subset of source-domain samples. The transferred knowledge mainly consists of the source-domain mean signal $\bar{\phi}^{(s)}$, the source-domain POD basis $\mathbf{M}_r^{(s)}$, normalization settings, and the pretrained FNN parameters. In Fig. 3, $W$ and $b$ denote the weight matrices and bias vectors of the pretrained FNN, respectively. Because the source and target domains share the same variables and the same phase-signal representation, we retain the reduced-order formulation introduced in Sections 2.1 and 2.2 during transfer. Specifically, each target-domain phase signal is projected onto the source-domain reduced basis according to $a_t(\mu) = \left(\mathbf{M}_r^{(s)}\right)^T \left(\phi_t(\mu) - \bar{\phi}^{(s)}\right)$, so that the learning task in the target domain remains the prediction of the reduced-order coefficients $a_t(\mu)$ from the input parameter vector $\mu$. To keep the target-domain data consistent with the pretrained surrogate, we apply the same normalization settings used in the source domain to both the target-domain inputs and the corresponding reduced-order coefficients.

To improve adaptation to the expanded parameter range while retaining the information already learned in the source domain, fine-tuning is carried out on a mixed dataset composed of all target-domain samples together with a retained subset of source-domain samples. Denoting the source-domain dataset and target-domain dataset by $D_s$ and $D_t$, respectively, the fine-tuning set is written as

$$D_{ft} = D_t \cup S(D_s, \rho), \tag{11}$$

where $S(D_s, \rho)$ denotes a randomly selected subset of the source-domain samples and $\rho$ is the retained fraction. In this work, $\rho = 0.1$ was adopted as a practical retained fraction to balance target-domain adaptation and source-domain knowledge retention. A representative comparison of phase-signal prediction accuracy obtained with different $\rho$ values is provided in the Supplementary Material. Based on this comparison, $\rho = 0.1$ was used in the following transfer-learning analyses. This mixed-data strategy allows the model to incorporate the newly generated target-domain samples while still preserving part of the source-domain information during fine-tuning.

After fine-tuning, the updated network predicts the reduced-order coefficients $\hat{a}_t(\mu)$ for target-domain samples, and the corresponding phase signal is reconstructed as $\hat{\phi}_t(\mu) = \bar{\phi}^{(s)} + \mathbf{M}_r^{(s)} \hat{a}_t(\mu)$. In this way, the POD-FNN surrogate can be extended to a broader parameter space with limited additional high-fidelity data while maintaining accurate forward prediction in the expanded domain.

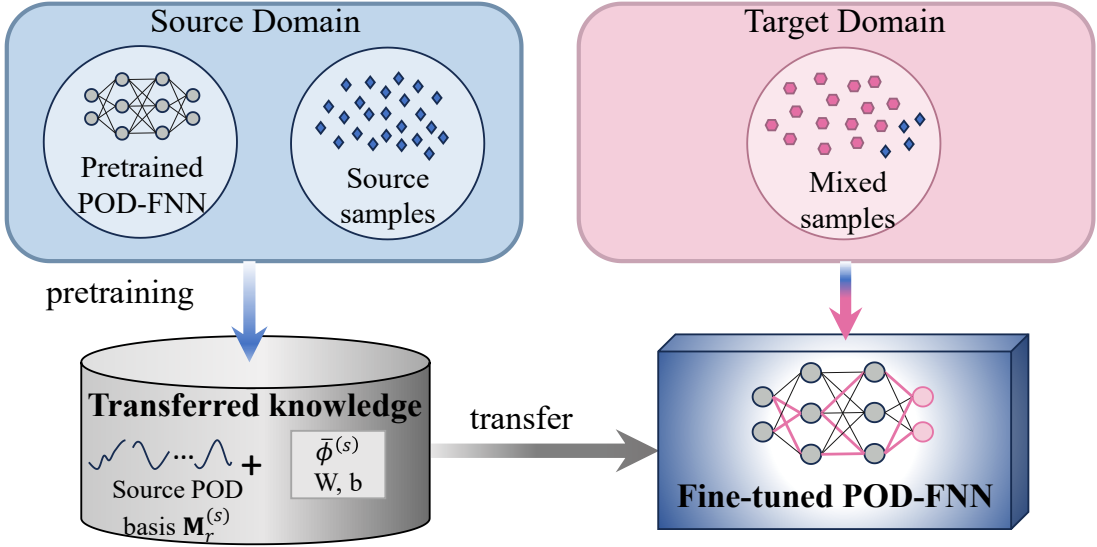

**Fig. 3.** Schematic illustration of the transfer-learning strategy for adapting the POD-FNN surrogate from the source domain to the target domain.

## 2.4 Overall framework

The overall framework of the proposed method is summarized in Fig. 4. The procedure consists of three main stages: high-fidelity data generation, reduced-order surrogate construction, and transfer adaptation. First, the prescribed thermophysical, geometric, and measurement parameters define the input parameter space, and the corresponding patterned thermoreflectance signals are generated by the COMSOL-based forward model. These phase signals are then organized into source-domain and target-domain datasets.

Next, POD is applied to the high-fidelity phase signals to express each response in terms of a small number of reduced-order coefficients. Based on this reduced-order representation, an FNN surrogate is trained to learn the mapping from the input parameter vector to the corresponding coefficients, from which the phase signal is reconstructed through the POD basis. When the parameter domain is expanded, additional target-domain samples are introduced and transfer learning is used to fine-tune the pretrained POD-FNN surrogate. In this way, the framework combines the accuracy of the high-fidelity solver, the efficiency of reduced-order surrogate modeling, and the adaptability provided by transfer learning.

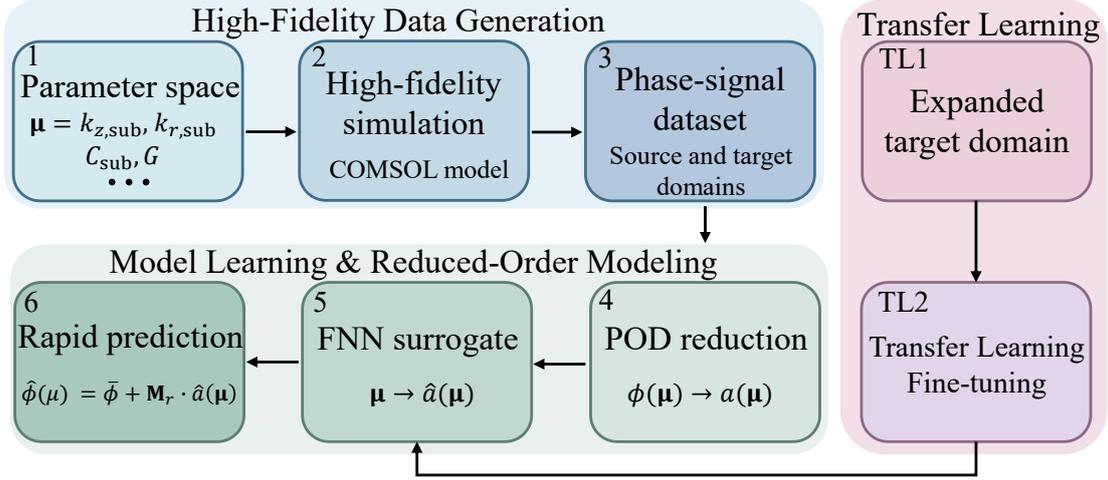

**Fig. 4.** Overall workflow of the proposed POD-FNN-TL framework.

## III. Experiment and Simulation Settings

### 3.1 Patterned FDTR

Conventional FDTR usually treats the transducer layer and the underlying sample as laterally semi-infinite. Under this condition, part of the deposited heat spreads radially in the upper region before it penetrates deeper into the material. Patterned FDTR modifies this heat-flow path by introducing a laterally confined surface structure whose characteristic radius is comparable to the pump size. Once the upper transducer region or the near-surface material becomes laterally confined, radial heat spreading weakens and a larger fraction of the thermal wave travels into the underlying layers. This geometric confinement changes the phase response and can improve sensitivity to selected thermal parameters, especially in structures where conventional semi-infinite configurations do not represent the actual heat-flow condition well. In this study, FDTR measurements were performed using a pump-probe optical configuration. A 458 nm continuous-wave pump laser was focused onto the sample surface to generate periodic heating, while a co-aligned 785 nm continuous-wave probe laser monitored the resulting thermoreflectance response. For phase-sensitive detection, a portion of the modulated pump beam was sampled as a reference signal and directed to a photodetector, with its optical path adjusted to match that of the reflected probe beam. The frequency-dependent phase signal recorded by the lock-in amplifier was subsequently used for thermal-model fitting and POD-FNN-based inverse analysis. A wider simulation frequency range was adopted during dataset generation to preserve model flexibility for broader forward-prediction tasks, whereas the experimental inversion was restricted to 10 kHz–10 MHz because this window provides adequate sensitivity to the substrate properties targeted in the present measurements and avoids unnecessary frequency regions with limited experimental benefit. The detailed optical layout has been reported previously in Ref. [24, 27], and only the key features relevant to the present work are summarized here.

In this work, FDTR serves as the experimental implementation of a broader patterned thermoreflectance framework. We focus on the frequency-dependent phase signal generated by a patterned metal transducer and use it as the target response for numerical simulation, surrogate construction, and subsequent parameter inversion. The geometric distinction between conventional and patterned configurations, together with the corresponding heat-flow features, will be discussed further in the following subsection.

### 3.2 Numerical simulations

We built the forward model in COMSOL Multiphysics 6.3 and controlled the entire model construction and solution process through MATLAB scripts via LiveLink for MATLAB. The implementation code is provided in the Supplementary Material. This approach allowed us to flexibly modify the model parameters and generate high-fidelity patterned FDTR signals for subsequent dataset construction. The numerical model follows a two-dimensional axisymmetric $r$-$z$ formulation, which matches the circular pump–probe geometry and the patterned transducer used in this work.

Fig. 5 compares the conventional FDTR model with the patterned FDTR model adopted here. In the conventional configuration, the transducer and substrate are treated as laterally semi-infinite. In contrast, the patterned configuration restricts the metal transducer to a finite radius $R_{\text{pattern}}$, while the substrate remains laterally extensive. This geometric modification changes the near-surface heat-flow path and suppresses lateral heat spreading in the transducer region, which is one of the main physical motivations for using patterned thermoreflectance measurements.

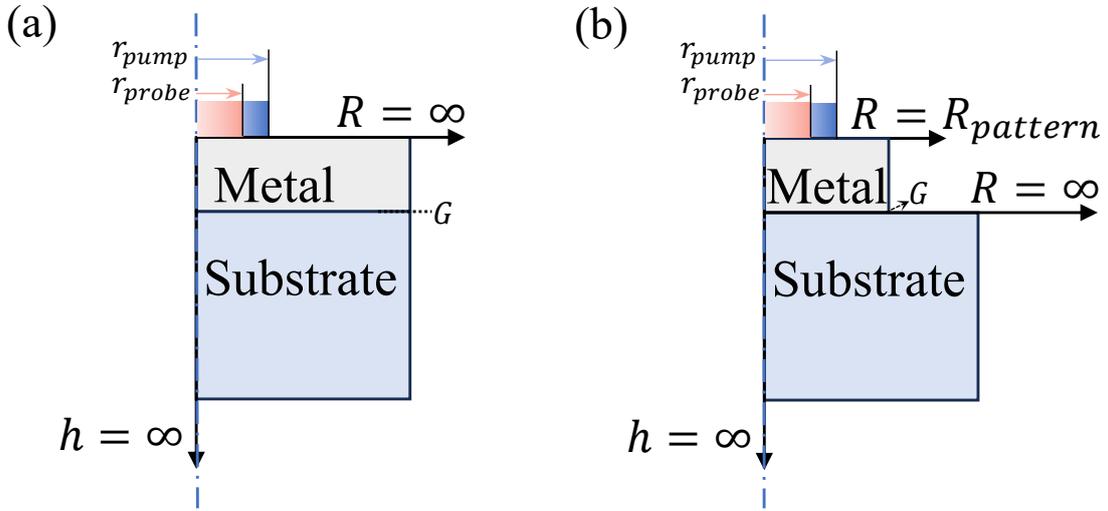

**Fig. 5.** Schematic comparison of (a) the conventional FDTR model and (b) the patterned FDTR model.

Based on this geometry, we constructed a 2D axisymmetric COMSOL model consisting of a metal transducer layer on top of a substrate domain, as illustrated in Fig. 6(a). We imposed a Gaussian harmonic heat flux on the top surface of the transducer to represent the modulated pump heating,

$$q''(r) = P \frac{2}{\pi r_{pump}^2} \exp\left(-\frac{2r^2}{r_{pump}^2}\right) \tag{12}$$

where $P$ is the pump power and $r_{pump}$ is the pump radius. We implemented the transducer/substrate interface beneath the patterned region in COMSOL using a thermal contact boundary, with the interfacial thermal resistance defined as $TBR = 1/G$, where $G$ is the interfacial conductance. Outside the patterned region, the exposed top surface of the substrate and the transducer sidewall were treated as insulation boundaries. To approximate the semi-infinite substrate condition within a finite computational domain, we fixed the far-right and bottom boundaries at $T = 0\ (K)$. The axis $r = 0$ satisfied the axisymmetric condition of the model. These settings are consistent with the model construction implemented in the MATLAB-controlled COMSOL workflow.

COMSOL solved the frequency-domain heat-transfer problem directly through its linearized harmonic formulation. In this framework, the model returns the complex temperature response associated with each modulation frequency rather than a time-domain transient temperature trace. We therefore evaluated the FDTR signal from the complex surface temperature on the transducer, instead of from a single nodal value. Specifically, we extracted the complex temperature along the top transducer surface and then calculated a probe-weighted average using the Gaussian probe profile with radius $r_{probe}$. The final phase signal was obtained from the argument of the weighted complex temperature. This treatment keeps the numerical post-processing consistent with the optical averaging process in FDTR measurements.

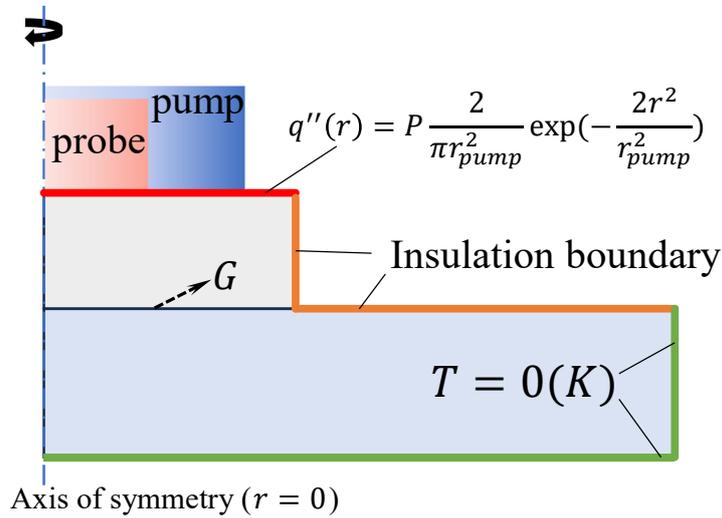

**Fig. 6.** Schematic of the 2D axisymmetric COMSOL model used for patterned FDTR simulations. A Gaussian harmonic heat flux is imposed on the top surface of the metal transducer. The exposed top surface of the substrate outside the patterned region and the transducer sidewall are treated as adiabatic, while the far-field boundaries are fixed at T=0 (K).

Fig. 7(a) shows a typical mesh used in the numerical model. The mesh is refined near the transducer edges and interface regions, where the temperature gradient varies most rapidly, and gradually coarsens away from the near-surface interaction zone. Fig. 7(b) shows a cross-sectional view of the magnitude of the harmonic temperature rise inside the sample at a modulation frequency of 10 kHz. The mesh used in this work was selected to balance numerical accuracy and computational efficiency over the frequency range of interest. A comparison of phase-signal predictions obtained with different mesh densities is provided in the Supplementary Material, showing that further mesh refinement leads to only minor changes in the predicted phase response within the present frequency window. For simulations at higher frequencies, a finer mesh may still be required to maintain the same numerical accuracy.

The COMSOL model described above serves as the high-fidelity forward solver in this work. By adjusting the model parameters in MATLAB and calling COMSOL to solve the corresponding frequency-domain problem, we generated the phase signals required for training dataset construction, surrogate development, and subsequent inverse analysis.

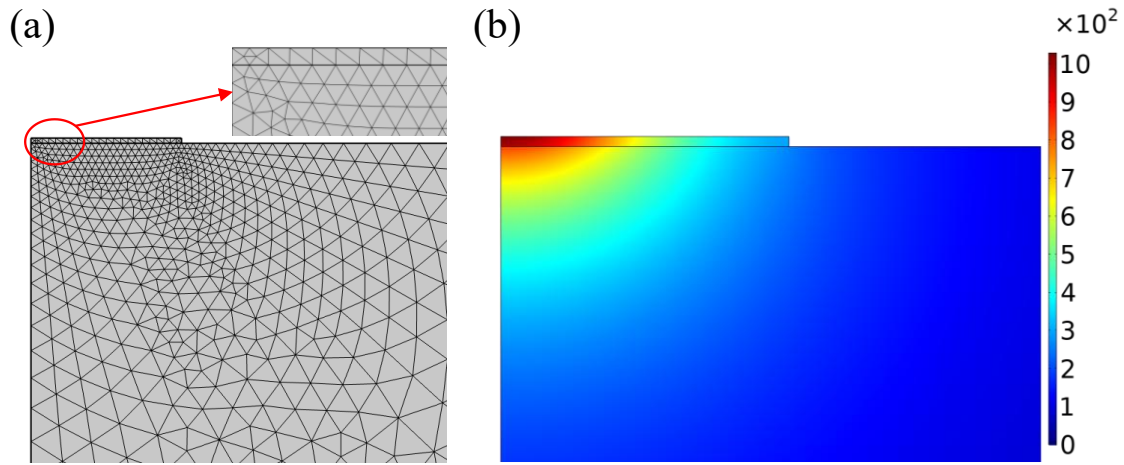

**Fig. 7.** (a) Representative mesh used in the numerical model, showing local refinement near the transducer edge and interface region; the inset provides an enlarged view of the refined mesh. (b) Cross-sectional view of the magnitude of the harmonic temperature rise within the sample at a modulation frequency of 10 kHz.

### 3.3 Training dataset settings

We generated the datasets by using Latin hypercube sampling with a fixed random seed and repeatedly calling the COMSOL forward solver through MATLAB scripts. For each sampled parameter set, the solver returned the patterned FDTR response over 100 logarithmically spaced modulation frequencies from 100 Hz to 30 MHz. During sampling, we varied the selected thermophysical, geometric, and optical parameters within prescribed ranges, while keeping the remaining values fixed. The sampled parameters and their ranges are summarized in Table 1. During dataset generation, all layers except the substrate were treated as isotropic, while the substrate thermal conductivity was allowed to vary anisotropically, with $k_{r,sub}$ and $k_{z,sub}$ sampled independently within prescribed ranges.

In this work, the metal transducer was taken as Al. Therefore, its volumetric heat capacity was kept fixed during dataset generation, while the transducer thickness was still allowed to vary. This treatment reflects the present material setting and avoids introducing an additional variable that is not central to the current study. When needed, however, the same framework can readily accommodate variations in the transducer heat capacity as well.

**Table 1.** Parameter ranges used for source- and target-domain dataset generation.

| Layer | Parameter | Unit | S domain N=6000 | FR domain N=1000 | OS domain N=1000 |
|---|---|---|---|---|---|
| Laser | $r_{pump}$ | μm | 0.8-2 | 0.8-2 | 0.8-2 |
|  | $r_{probe}$ | μm | 0.8-2 | 0.8-2 | 0.8-2 |
| Al(transducer) | $k_{r,m}$ | W/(m·K) | 30-200 | **10-200** | **10-50** |
|  | $k_{z,m}$ | W/(m·K) | 30-200 | **10-200** | **10-50** |
|  | $C_m$ | MJ/(m³·K) | 2.2-2.6 | 2.2-2.6 | 2.2-2.6 |
|  | $h_m$ | nm | 50-150 | 50-150 | 50-150 |
|  | $R_{Pat}$ | μm | 1.5-5 | 1.5-5 | 1.5-5 |
| interface | $G$ | MW/(m²·K) | 50-200 | 50-200 | 50-200 |
| substrate | $k_{r,sub}$ | W/(m·K) | 0.1-10 | 0.1-10 | 0.1-10 |
| | $k_{z,sub}$ | W/(m·K) | 0.1-10 | **0.1-100** | **5-100** |
| | $C_{sub}$ | MJ/(m³·K) | 0.3-3 | 0.3-3 | 0.3-3 |
| model domain | $R_{max}$ | μm | 30 | 30 | 30 |

To support surrogate training and transfer-learning evaluation, three training datasets were generated. The 6000-sample dataset defines the source domain and was used to train the baseline POD-FNN surrogate in the original parameter space. Two target domain datasets were further constructed for domain-expansion studies. One is a full-range target dataset (FR), which covers the entire extended parameter domain, including both the source and extended domains. The other is an out-of-source-focused dataset (OS), which samples only the region outside the original source domain and therefore represents a more concentrated expansion scenario. The expanded domains were designed to mimic a practical model-updating scenario in which the surrogate trained in an initial material range must later be adapted to transducer and substrates with substantially broader conductivity while keeping the remaining experimental and geometric settings in a realistic range. Their detailed parameter ranges are listed in Table 1.

### 3.4 Sample preparation

Patterned $Al/SiO_2$ samples were fabricated on 3-inch SiO2 substrates using standard photolithography and lift-off techniques[28]. A 4-inch photomask containing the designed patterns was first employed to define the patterned regions on the substrate surface. After lithographic development, an Al film was deposited on the substrate by electron-beam evaporation. The sample was subsequently immersed in acetone for 30 min and then treated in an ultrasonic bath for 2 min to assist the lift-off process. This process removed the photoresist together with the undesired Al layer, leaving the patterned Al film on the SiO2 substrate. A representative optical micrograph of the fabricated sample is shown in Fig. 8, confirming that the designed pattern was successfully formed after the lift-off process. In addition, supplementary material shows that, within the experimental frequency range of 10 kHz–10 MHz, the selected outer radius $R$max is sufficiently large and the corresponding finite-size effect is negligible. Therefore, the simplified geometry adopted in the COMSOL model is justified for the present study.

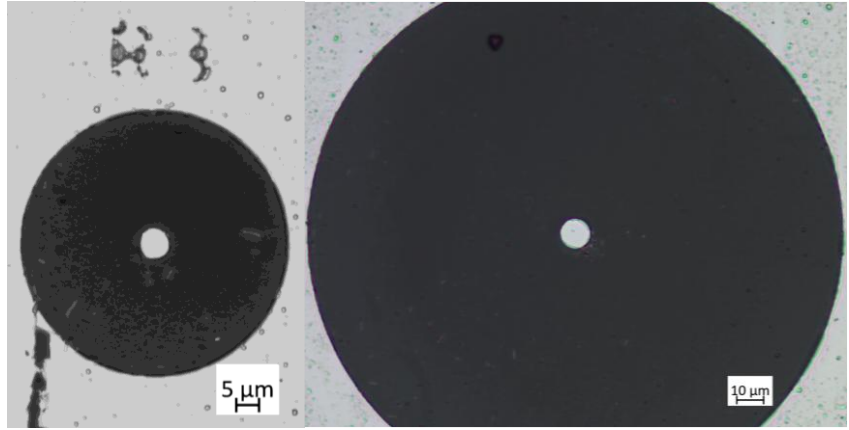

**Fig. 8.** 50X Optical micrograph of patterned $Al/SiO_2$ sample

## IV. Results and Discussion

### 4.1 Validation of the high-fidelity forward model

The accuracy of the developed COMSOL model was assessed by comparing the predicted FDTR phase signals with the corresponding results extracted from the literature under identical material properties and geometric conditions. As shown in Fig. 9, the simulated phase signals agree well with the digitized literature data throughout the investigated frequency range.

The agreement holds for both the conventional FDTR configuration and the patterned FDTR configurations with different disk radii. Moreover, the present model successfully captures the characteristic frequency-dependent differences among the various structures, including the distinct phase separation at low and intermediate frequencies and the gradual convergence of the curves at higher frequencies. The phase variation associated with changes in patterned radius is also reproduced accurately, further demonstrating that the model correctly reflects the effect of structural size on the thermal response.

Overall, the comparison verifies that the developed COMSOL model provides an accurate description of the FDTR response reported in the literature. The model was therefore used as the high-fidelity forward model for subsequent dataset generation, POD-based dimensionality reduction, surrogate model construction, and parameter fitting. In addition to the literature-based forward validation shown here, the applicability of the model to practical signal analysis is further supported in Section 4.3 through surrogate-assisted fitting of measured FDTR data.

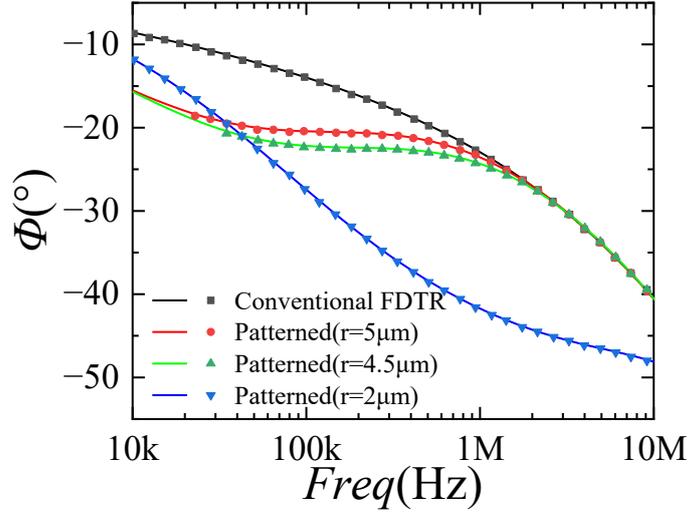

**Fig. 9.** Validation of the COMSOL model by FDTR phase signal comparison for conventional and patterned structures with different disk radii. Solid lines represent the present COMSOL results, and symbols denote the digitized literature data[8].

### 4.2 Accuracy and efficiency of the POD-FNN surrogate

To evaluate the predictive performance of the POD-FNN surrogate within the original parameter domain, an independent test set consisting of 100 randomly generated signals was constructed. For the $i$-th test sample with parameter vector $\mu^{(i)}$, the phase signal predicted by the surrogate, $\hat{\phi}(\mu^{(i)}, f_j)$, was compared with the corresponding high-fidelity COMSOL result, $\phi(\mu^{(i)}, f_j)$, at the $j$-th frequency point $f_j$. The prediction error for each test sample was quantified by the per-sample root mean square error (RMSE), defined as

$$\mathrm{RMSE}_i = \sqrt{\frac{1}{N_f}\sum_{j=1}^{N_f}[\hat{\phi}(\mu^{(i)},f_j)-\phi(\mu^{(i)},f_j)]^2}, \tag{13}$$

where $N_f$ is the number of frequency points. In addition, the frequency-dependent prediction error over the full test set was characterized by the mean absolute error (MAE),

$$\mathrm{MAE}(f_j) = \frac{1}{N_{\text{test}}}\sum_{i=1}^{N_{\text{test}}}|\hat{\phi}(\mu^{(i)},f_j)-\phi(\mu^{(i)},f_j)|, \tag{14}$$

where $N_{\text{test}}$ is the number of test samples.

Representative comparisons for the best, median, and worst cases are shown in Fig. 10(a). In all three cases, the POD-FNN-predicted signals closely follow the COMSOL results over the entire investigated frequency range, indicating that the surrogate is able to accurately reproduce the frequency-dependent phase response of the patterned FDTR model.

The overall prediction accuracy of the surrogate on the independent test set is further summarized in Fig. 10(b) and Fig. 10(c). As shown in Fig. 10(b), the per-sample RMSE values are mainly distributed in the low-error region. The mean and median RMSE are 0.19° and 0.17°, respectively, while the maximum RMSE remains below 0.47°. In addition, 87% of the test samples have RMSE values below 0.3°, further indicating that the surrogate maintains low prediction error for most cases in the original parameter domain. Figure 10(c) presents the frequency-dependent MAE, which remains at a relatively low level throughout the full frequency range considered. Although the error increases gradually from the low-frequency region to the intermediate- and high-frequency ranges, its overall magnitude remains small, demonstrating that the surrogate maintains stable predictive accuracy over the frequency domain of interest. The gradual increase in MAE toward the intermediate- and high-frequency range is likely related to the stronger influence of near-surface parameters under shallower thermal penetration conditions, which makes the phase response more sensitive to local variations in transducer- and interface-related quantities.

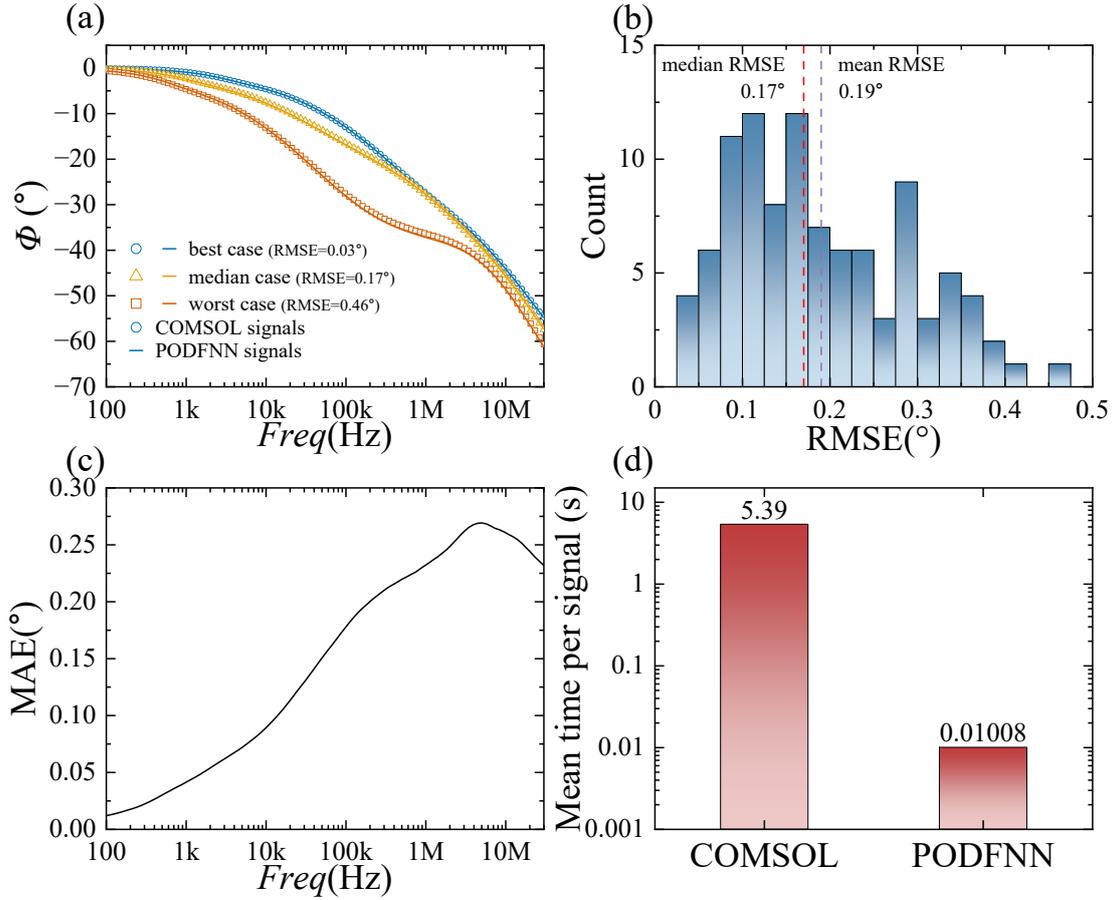

**Fig. 10.** Performance evaluation of the POD-FNN surrogate on an independent test set of 100 randomly generated signals within the original parameter domain: (a) representative comparisons of COMSOL and POD-FNN signals for the best, median, and worst cases; (b) distribution of per-sample root mean square error (RMSE); (c) frequency-dependent mean absolute error (MAE); and (d) comparison of average prediction time per sample for COMSOL and POD-FNN.

The computational efficiency of the POD-FNN surrogate is compared with that of the full-order COMSOL model in Fig. 10(d). The average time required for a single signal prediction is reduced from 5.39 s for the COMSOL model to 0.0101 s for the POD-FNN surrogate, corresponding to a speedup of approximately 534×. These results show that the POD-FNN surrogate achieves high prediction accuracy together with substantial computational acceleration within the original parameter domain. Its advantage becomes particularly evident when repeated forward evaluations are required in iterative fitting and related analyses, and it also provides a suitable basis for subsequent transfer-learning-based extension.

### 4.3 Surrogate-based inverse analysis and parameter fitting

In the inverse analysis, a hybrid optimization strategy combining particle swarm optimization (PSO) and the BFGS algorithm[27] was adopted. PSO was first used for global exploration within the prescribed parameter bounds, and the resulting solution was then refined by BFGS for local optimization. In the present work, the PSO swarm size was set to 200 and the maximum number of PSO iterations was set to 20. Before performing the large-scale inverse validation, a representative synthetic fitting case was first examined in which the interfacial thermal conductance $G$, the substrate in-plane thermal conductivity $k_{r,sub}$, and the substrate cross-plane thermal conductivity $k_{z,sub}$ were simultaneously treated as fitting parameters. The target phase signal was generated by the full-order COMSOL model, and the recovered parameter values obtained from POD-FNN-based fitting and direct COMSOL-based fitting are compared in Table 2. For fairness, the direct COMSOL fitting and the POD-FNN fitting were carried out under the same PSO optimization settings. As shown in Table 2, for this representative case, the POD-FNN fitting produced a slightly lower phase RMSE than the direct COMSOL fitting in this representative example, while maintaining comparable recovery accuracy for the substrate thermal conductivities at a much lower computational cost. Therefore, the surrogate-based inversion was considered sufficiently accurate for the subsequent large-scale validation and measured-data fitting tasks. Although both approaches reproduced the target phase signal with low fitting error, the recovered value of $G$ showed a noticeably larger deviation from the prescribed value than those of $k_{r,sub}$ and $k_{z,sub}$. This result suggests that, under the present frequency window and sensitivity conditions, simultaneous estimation of $G$, $k_{r,sub}$, and $k_{z,sub}$ is more challenging than inversion of the substrate thermal conductivities alone. Accordingly, in the following statistical inverse validation, $G$ was fixed at its prescribed value and only $k_{r,sub}$ and $k_{z,sub}$ were treated as fitting parameters.

Table 2. Comparison of prescribed and recovered parameters for a representative synthetic inverse-fitting case.

| Method | $G$ (MW/(m² · K)) | $k_{r,sub}$ W/(m · K) | $k_{z,sub}$ W/(m · K) | Phase RMSE (°) | Time(s) |
|---|---|---|---|---|---|
| Prescribed | 100 | 1.38 | 1.38 | --- | --- |
| POD-FNN fitting | 90.3 | 1.370 | 1.404 | 0.101 | 65 |
| COMSOL fitting | 83.5 | 1.368 | 1.376 | 0.135 | 18950 |

Following the representative three-parameter fitting test described above, the large-scale inverse validation was carried out with the interfacial thermal conductance fixed at its prescribed value. Under this setting, 200 synthetic phase signals generated by the full-order COMSOL model were fitted using the POD-FNN inversion framework. The substrate in-plane and cross-plane thermal conductivities, $k_{r,sub}$ and $k_{z,sub}$, were treated as the only fitting parameters, while the remaining quantities were fixed at their nominal values. The corresponding fixed settings are provided in the Supplementary Material.

Figure 11 presents the comparisons between the prescribed parameter values used in the COMSOL simulations and the corresponding values identified by the surrogate-based inversion. In each panel, the dashed line denotes the ideal relation $y = x$, and the two solid lines represent the $\pm 5\%$ deviation bands. Specifically, 88% of the fitted $k_{r,sub}$ values fall within the $\pm 5\%$ bounds, while the fitted $k_{z,sub}$ values also remain largely clustered around the $y = x$ line. As shown in Fig. 11(a) and Fig. 11(b), the fitted values of $k_{r,sub}$ and $k_{z,sub}$ remain in close correspondence with the true values over the investigated parameter range, with the majority of the data points falling within the $\pm 5\%$ bounds. These results indicate that the POD-FNN surrogate is capable of reproducing the inverse solutions of the full-order COMSOL model with satisfactory accuracy, while substantially reducing the need for repeated high-fidelity forward evaluations during the iterative fitting process. For a representative inverse-fitting case under the same optimization settings, the POD-FNN framework reduced the fitting time from about 18950 s for the COMSOL-based optimization to about 65 s, while maintaining comparable fitting accuracy.

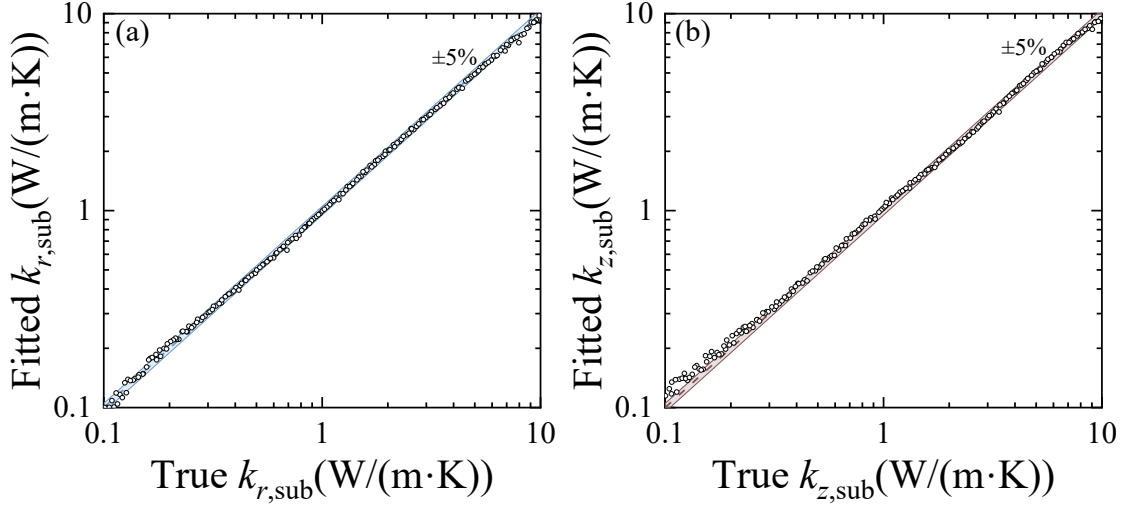

**Fig. 11.** Inverse validation of the POD-FNN surrogate using 200 COMSOL-generated synthetic signals: comparisons between the true and fitted values of (a) $k_{r,sub}$, (b) $k_{z,sub}$. The dashed line represents $y = x$, and the two solid lines indicate the $\pm 5\%$ deviation from the $y = x$ line.

Following the inverse validation on synthetic signals, the proposed framework was further applied to measured FDTR data. In the fitting of the measured signals, the thermal conductivity of the Al transducer was also treated as an adjustable parameter. Because the patterned Al transducer was fabricated through the lift-off process, its fitted thermal conductivity may differ from that of the Al film used in the conventional FDTR measurement[29]. Allowing this parameter to vary therefore helps reduce possible model mismatch in the inverse analysis. This treatment also makes the present experimental inversion more conservative, since uncertainty associated with the patterned transducer is not neglected a priori. The interfacial thermal conductance was not treated as a fitting parameter in the experimental inversion. As shown in Fig. 12(b), under the present measurement conditions and frequency window, the phase sensitivity to the interfacial conductance remained much lower than that to the substrate thermal conductivities and stayed close to zero over most of the investigated frequency range, making reliable simultaneous estimation of this parameter difficult. Therefore, the interfacial conductance was fixed at the prescribed value, and its uncertainty was included in the subsequent uncertainty analysis. The fitted thermal conductivities of the Al transducer and silica substrate for the three measurement configurations are summarized in Table 3. The uncertainties reported in Table 3 were estimated by sensitivity-based first-order covariance propagation, in which the prescribed uncertainties of the fixed input quantities, including the interfacial conductance, were propagated through the sensitivity matrix evaluated at the fitted solution; the reported values correspond to 2σ uncertainty bounds. Figure 12(a) shows the measured phase signals together with the corresponding POD-FNN-based fitting results for conventional FDTR and patterned FDTR with two pattern radii, $R_{pat} = 5.3$ μm and $3.25$ μm. For all three cases, the fitted curves reproduce the overall frequency-dependent trends of the measured signals over the investigated frequency range, demonstrating the capability of the surrogate-based framework to capture the experimental phase response. The fitted Al transducer conductivity remains within a relatively narrow range of 44–47 W/(m·K) across the three measurement configurations, indicating that the fitted transducer parameter is reasonably consistent among different datasets. More importantly, the extracted silica thermal conductivity is also stable, with values of $1.44 \pm 0.088$, $1.43 \pm 0.093$, and $1.50 \pm 0.079$ W/(m·K) for conventional FDTR, patterned FDTR with $R_{pat} = 5.3$ μm, and patterned FDTR with $R_{pat} = 3.25$ μm, respectively.

Table 3. Fitted thermal conductivities of the silica substrate and Al transducer obtained from measured FDTR signals under different transducer configurations.

| Configuration | Fitted $k_{\text{Silica}}$ W/(m·K) | Fitted $k_{\text{Al}}$ W/(m·K) |
|---|---|---|
| Conventional FDTR | $1.44 \pm 0.088$ | $47 \pm 2.78$ |
| $R_{pat} = 5.3$ μm | $1.43 \pm 0.093$ | $44 \pm 1.98$ |
| $R_{pat} = 3.25$ μm | 1.50 ± 0.079 | 47 ± 2.41 |

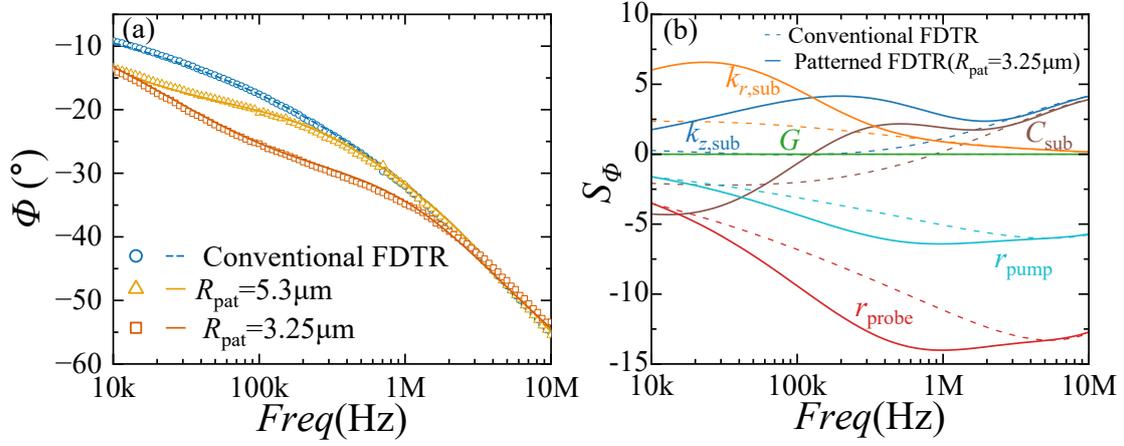

**Fig. 12.** Experimental phase signals, surrogate-based fitting results, and representative phase sensitivities for conventional and patterned FDTR: (a) measured phase signals and corresponding POD-FNN fitting curves for conventional FDTR and patterned FDTR with $R_{pat} = 5.3$ μm and 3.25 μm, symbols denote measured data, and lines denote POD-FNN fitting curves; (b) phase sensitivities $S_\phi$ of selected parameters, including $k_{z,\text{sub}}$, $k_{r,\text{sub}}$, $C_{\text{sub}}$, $G$, $r_{pump}$, $r_{probe}$, for patterned FDTR with $R_{pat} = 3.25$ μm (solid lines) and conventional FDTR (dashed lines).

To provide physical interpretation for the fitting performance, the phase sensitivities of selected parameters are compared in Fig. 12(b). The patterned FDTR case with $R_{pat} = 3.25$ μm is presented as a representative example, and the corresponding sensitivities for conventional FDTR are included for reference. In comparison with the conventional configuration, the patterned transducer modifies the sensitivity distribution and increases the sensitivity of the phase signal to the substrate thermal conductivities, particularly to the in-plane and cross-plane components.

Taken together, these results demonstrate that the POD-FNN surrogate provides a reliable approximation not only for forward prediction but also for inverse parameter estimation. Under the present inversion setting, in which the transducer thermal conductivity was also allowed to vary to reduce possible model mismatch, the proposed framework enables stable substrate-property extraction for low-thermal-conductivity materials while substantially reducing the computational cost associated with repeated high-fidelity evaluations. From a practical standpoint, this computational gain is important not only for single-point parameter extraction, but also for characterization scenarios involving repeated measurements at multiple locations. In such cases, the reduced fitting time can facilitate high-throughput property mapping and more efficient assessment of spatial non-uniformity in thin films and interfaces. Therefore, the present POD-FNN framework is not merely a surrogate for numerical acceleration, but a potentially useful computational component for rapid thermophysical characterization workflows.

### 4.4 Transfer learning for expanded parameter domains

When the parameter domain is enlarged beyond the original source range, the predictive accuracy of the baseline surrogate deteriorates noticeably. To improve the applicability of the POD-FNN surrogate in the broadened parameter space, transfer learning was introduced using two types of target-domain datasets, namely a full-range expanded target dataset (FR) and an out-of-source-focused target dataset (OS). The specific parameter ranges are listed in Table 1.

Representative predictions are compared in Fig. 13(a) and Fig. 13(b). For both test samples, the source-domain POD-FNN model deviates from the COMSOL reference after the parameter domain is expanded, and the discrepancy becomes particularly pronounced for the more challenging case in Fig. 13(b). By contrast, both TL-FR and TL-OS recover the phase response much more accurately over the whole frequency range. These results indicate that transfer learning enables more accurate adaptation of the surrogate to the expanded parameter domain.

A quantitative comparison is provided in Fig. 13(c) through the distribution of per-sample RMSE on an independent test set containing 200 samples. Relative to the source-domain POD-FNN, all adapted models markedly reduce the prediction error in the expanded domain. Moreover, under both FR and OS data settings, the transfer-learned surrogates exhibit lower RMSE distributions than the corresponding directly trained POD-FNN models. The median RMSE values of the source-domain POD-FNN, direct POD-FNN-FR, TL-FR, direct POD-FNN-OS, and TL-OS are 0.578°, 0.512°, 0.210°, 1.364°, and 0.362°, respectively. Under the present training and test setup, TL-FR achieves the lowest median RMSE and therefore gives the best overall prediction performance in the expanded parameter domain. This result suggests that, when only a limited number of additional high-fidelity samples are available, transfer learning can make more efficient use of the source-domain knowledge and achieve more accurate adaptation to the expanded parameter domain. In addition, the use of reduced target-domain training sets substantially lowers the cost of high-fidelity data generation. For the present cases, generating 6000 samples required about 34179 s, whereas generating 1000 samples required only about 5885 s, corresponding to a reduction of approximately 82.8%. By comparison, the neural-network training time remained of the same order, indicating that the main cost saving arises from the reduced need for additional high-fidelity simulations.

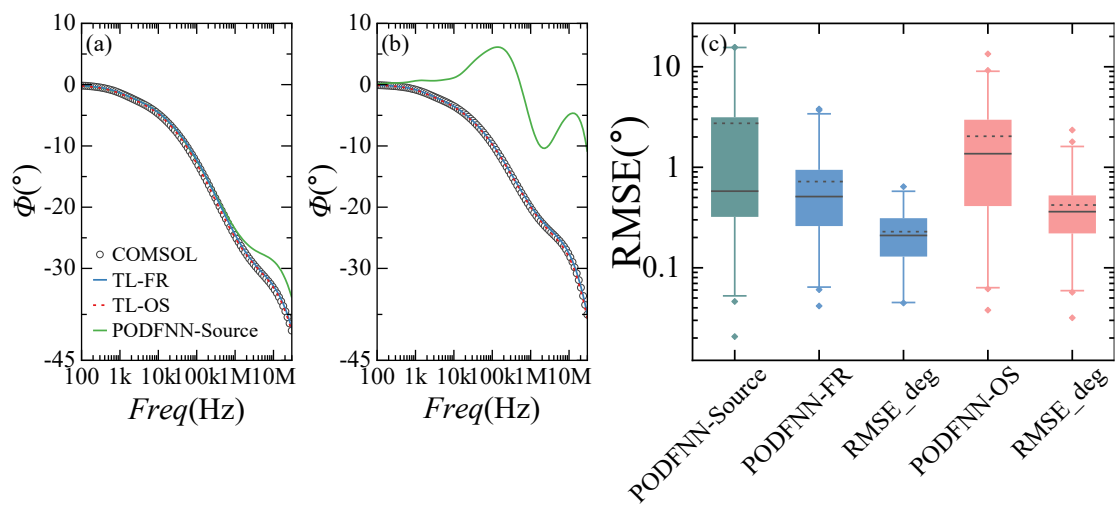

**Fig. 13.** Transfer-learning performance of the POD-FNN surrogate in the expanded parameter domain. (a, b) Representative phase-signal predictions for two samples from the expanded parameter domain, comparing COMSOL results with source-domain POD-FNN and the transfer-learned surrogates trained using the full-range target dataset (FR) and the out-of-source-focused target dataset (OS). (c) Distribution of per-sample RMSE on an independent test set ($N_{test} = 200$) for source-domain POD-FNN, directly trained POD-FNN, and transfer-learned POD-FNN.

Among the compared models, TL-FR showed the best overall performance under the present setup. Since the FR dataset covers both the original source region and the expanded parameter region, it provides a more balanced set of samples for fine-tuning. This broader coverage helps the pretrained surrogate retain source-domain knowledge while adapting to the expanded domain, which likely contributes to the lower RMSE distribution achieved by TL-FR. Overall, these results demonstrate that transfer learning provides an effective route for extending the POD-FNN surrogate to broader parameter domains while limiting the amount of additional high-fidelity training data required. From an engineering perspective, this transferability is valuable because practical thermoreflectance workflows often need to accommodate changes in sample structure, film quality, interface condition, or measurement setup without rebuilding the surrogate from scratch. The present transfer-learning strategy therefore offers a practical route for maintaining rapid prediction and fitting capability as the relevant material or process window evolves.

## V. Conclusion

A transfer-learning-enhanced POD-FNN framework was developed for patterned thermoreflectance and validated using patterned FDTR as a representative case. Within the original parameter domain, the surrogate achieved mean and median RMSE values of 0.19° and 0.17°, respectively, with the maximum RMSE below 0.47°. Meanwhile, the average prediction time per signal was reduced from 5.39 s for COMSOL to 0.01 s, corresponding to a speedup of about 534×. In inverse analysis, the POD-FNN-based fitting reduced the optimization time for a representative case from about 18950 s to about 65 s while maintaining comparable accuracy. For measured $Al/SiO_2$ samples, the extracted silica thermal conductivity remained stable at $1.44 \pm 0.088$, $1.43 \pm 0.093$, and $1.50 \pm 0.079$ W/(m·K) for conventional FDTR and patterned FDTR with $R_{pat} = 5.3$ μm and 3.25 μm, respectively. After expansion of the parameter domain, transfer learning further improved the surrogate performance, with TL-FR yielding the lowest median RMSE among the compared models. In addition, reducing the additional target-domain dataset from 6000 to 1000 samples lowered the high-fidelity data-generation time from about 34179 s to about 5885 s. The present experimental validation was limited to one representative $Al/SiO_2$ material system. Broader experimental verification on additional low-thermal-conductivity materials would further strengthen the generality of the proposed framework. Overall, the proposed framework provides a practical route for patterned thermoreflectance workflows that require rapid forward evaluation, repeated inverse fitting, and efficient surrogate updating as the parameter domain expands. Beyond computational acceleration, the broader value of the present framework lies in supporting faster thermal-characterization workflows for thin films, buried interfaces, and other interface-dominated structures. Such capability is relevant to high-throughput property screening, spatial mapping, and model recalibration during thermal-material and device development.

**References**

[1] R. Mohan, S. Khan, R.B. Wilson, P.E. Hopkins, Time-domain thermoreflectance, Nature Reviews Methods Primers, 5 (2025) 55.

[2] P. Jiang, X. Qian, R. Yang, Tutorial: Time-domain thermoreflectance (TDTR) for thermal property characterization of bulk and thin film materials, Journal of Applied Physics, 124 (2018).

[3] T. Chen, S. Song, Y. Shen, K. Zhang, P. Jiang, Simultaneous measurement of thermal conductivity and heat capacity across diverse materials using the square-pulsed source (SPS) technique, International Communications in Heat and Mass Transfer, 158 (2024) 107849.

[4] D.J. Kirsch, J. Martin, R. Warzoha, M. McLean, D. Windover, I. Takeuchi, An instrumentation guide to measuring thermal conductivity using frequency domain thermoreflectance (FDTR), Review of Scientific Instruments, 95 (2024).

[5] L. Tang, C. Dames, Anisotropic thermal conductivity tensor measurements using beam-offset frequency domain thermoreflectance (BO-FDTR) for materials lacking in-plane symmetry, International Journal of Heat Mass Transfer, 164 (2021) 120600.

[6] P. Jiang, D. Wang, Z. Xiang, R. Yang, H. Ban, A new spatial-domain thermoreflectance method to measure a broad range of anisotropic in-plane thermal conductivity, International Journal of Heat Mass Transfer, 191 (2022) 122849.

[7] M. Goni, M. Patelka, S. Ikeda, T. Sato, A.J. Schmidt, Frequency domain thermoreflectance technique for measuring the thermal conductivity of individual

micro-particles, Review of Scientific Instruments, 89 (2018).
[8] Y. Akura, Y. Matsunaga, L. Liu, Y. Ikeda, M. Shimofuri, A. Banerjee, T. Tsuchiya, J. Hirotani, Enhancing the precision of thermal conductivity measurement via transducer patterning in frequency-domain thermoreflectance, Review of Scientific Instruments, 96 (2025).
[9] R.J. Warzoha, A.A. Wilson, B.F. Donovan, A.N. Smith, N. Vu, T. Perry, L. Li, N. Miljkovic, E. Getto, A numerical fitting routine for frequency-domain thermoreflectance measurements of nanoscale material systems having arbitrary geometries, Journal of Applied Physics, 129 (2021).
[10] R.J. Warzoha, A.A. Wilson, B.F. Donovan, A. Clark, X. Cheng, L. An, G. Feng, Measurements of thermal resistance across buried interfaces with frequency-domain thermoreflectance and microscale confinement, ACS Applied Materials Interfaces, 16 (2024) 41633-41641.
[11] X.-Y. Feng, F. Bai, W.-Q. Tao, A new efficient conservation-based method for implementing pod-Galerkin projection, in: International Heat Transfer Conference Digital Library, Begel House Inc., 2023.
[12] L. Xiang, B. Zhang, Y. Zha, G. Xing, X. Yang, Z. Wang, Y. Cheng, X. Yu, R. Hu, X. Luo, Physics-informed proper orthogonal decomposition for accurate and superfast prediction of thermal field, ASME Journal of Heat Mass Transfer, 147 (2025) 073301.
[13] M. Allabou, R. Bouclier, P.-A. Garambois, J. Monnier, Reduction of the shallow water system by an error aware POD-neural network method: Application to floodplain dynamics, Computer Methods in Applied Mechanics Engineering, 428 (2024) 117094.
[14] Y.-Q. Tang, W.-Z. Fang, C.-Y. Zheng, W.-Q. Tao, Applications of POD-based reduced order model to the rapid prediction of velocity and temperature in data centers, Applied Thermal Engineering, 263 (2025) 125310.
[15] X. Li, Q. Xu, S. Wang, K. Luo, J. Fan, A novel data-driven reduced-order model for the fast prediction of gas-solid heat transfer in fluidized beds, Applied Thermal Engineering, 253 (2024) 123670.
[16] L. Xiang, F. Wang, Y. Zha, Y. Hu, B. Zhang, X. Yang, R. Hu, X. Luo, Reduced-order-driven recurrent neural network for ultra-fast thermal field simulation in high-heat-flux electronic systems, International Journal of Heat Mass Transfer, 251 (2025) 127351.
[17] F. Wang, Y. Hu, B. Zhang, Y. Zha, X. Luo, Fast Temperature Field Extrapolation Under Non-Periodic Boundary Conditions, Applied Sciences, 15 (2025) 3895.
[18] S.W. Chung, Y. Choi, P. Roy, T. Moore, T. Roy, T.Y. Lin, D.T. Nguyen, C. Hahn, E.B. Duoss, S.E. Baker, Train small, model big: Scalable physics simulators via reduced order modeling and domain decomposition, Computer Methods in Applied Mechanics and Engineering, 427 (2024) 117041.
[19] R. Fu, D. Xiao, A.G. Buchan, X. Lin, Y. Feng, G. Dong, A parametric non-linear non-intrusive reduce-order model using deep transfer learning, Computer Methods in Applied Mechanics and Engineering, 438 (2025) 117807.
[20] A.H. Ali, G.Y. Mohanad, A. Mohammad, A. Saad Abbas, Transfer Learning: A

New Promising Techniques, Mesopotamian Journal of Big Data, 2023 (2023) 29-30.
[21] W. Guo, F. Zhuang, X. Zhang, Y. Tong, J. Dong, A comprehensive survey of federated transfer learning: challenges, methods and applications, Frontiers of Computer Science, 18 (2024) 186356.
[22] J. Gupta, S. Pathak, G. Kumar, Deep learning (CNN) and transfer learning: a review, in: Journal of Physics: Conference Series, Vol. 2273, IOP Publishing, 2022, pp. 012029.
[23] S. Zhang, C. Deng, F. Wei, J. Li, X. Luo, J. Ma, Efficient structure optimization of downhole thermal management system via transfer learning surrogate modeling, Journal of Energy Storage, 151 (2026) 120543.
[24] T. Chen, P. Jiang, Decoupling thermal properties in multilayered systems for advanced thermoreflectance experiments, Physical Review Applied, 23 (2025) 044004.
[25] N. Poopakdee, Z. Abdallah, J.W. Pomeroy, M. Kuball, In situ thermoreflectance characterization of thermal resistance in multilayer electronics packaging, ACS Applied Electronic Materials, 4 (2022) 1558-1566.
[26] D.B. Brown, W. Shen, X. Li, K. Xiao, D.B. Geohegan, S. Kumar, Spatial mapping of thermal boundary conductance at metal–molybdenum diselenide interfaces, ACS Applied Materials Interfaces, 11 (2019) 14418-14426.
[27] B. Xiao, T. Chen, W. Zhang, X. Qian, P. Jiang, Fast and reliable thermal property extraction from FDTR measurements via hybrid particle swarm optimization, International Journal of Heat Mass Transfer, 256 (2026) 128109.
[28] S.-K. Fan, C.-J. Kim, J.-A. Paik, B. Dunn, P.R. Patterson, M.C. Wu, MEMS with thin-film aerogel, in: Technical Digest. MEMS 2001. 14th IEEE International Conference on Micro Electro Mechanical Systems (Cat. No. 01CH37090), IEEE, 2001, pp. 122-125.
[29] T. Duy Hien, R.A. Zwijze, J.W. Berenschot, R.J. Wiegerink, G.J.M. Krijnen, M. Elwenspoek, Platinum Patterning by a Modified Lift-Off Technique and Its Application in a Silicon Load Cell, Sensors and Materials, 13 (2001).